\documentclass[twocolumn,preprintnumbers,amsmath,amssymb]{revtex4}
\usepackage{graphicx}
\usepackage{dcolumn}
\usepackage{bm}
\usepackage{times}
\usepackage[colorlinks,citecolor=blue,linkcolor=red]{hyperref}
\usepackage{color}
\usepackage{comment}

\begin{document}

\title{Negative differential thermal conductance and heat amplification in a nonequilibrium triangle-coupled spin-boson system at strong coupling}
\author{Chen Wang$^{1,2,}$}\email{wangchenyifang@gmail.com}
\author{Xu-Min Chen$^{1}$}
\author{Ke-Wei Sun$^{1}$}
\author{Jie Ren$^{3,}$}\email{Xonics@tongji.edu.cn}
\address{
$^{1}$Department of Physics, Hangzhou Dianzi University, Hangzhou 310018, Zhejiang, P. R. China\\
$^{2}$Department of Physics, Zhejiang Normal University, Jinhua 321004, Zhejiang , P. R. China\\
$^{3}$Center for Phononics and Thermal Energy Science, China-EU Joint Center for Nanophononics, \\
Shanghai Key Laboratory of Special Artificial Microstructure Materials and Technology,  \\
School of Physics Sciences and Engineering, Tongji University, Shanghai 200092, China
}

\date{\today}

\begin{abstract}
We investigate the nonequilibrium quantum heat transfer in a triangle-coupled spin-boson system within a three-terminal setup.
By including the nonequilibrium noninteracting blip approximation approach combined with the full counting statistics, we analytically obtain the steady state populations and heat currents.
The negative differential thermal conductance and giant heat amplification factor are clearly observed at strong qubit-bath coupling.
%and the heat amplification is dramatically suppressed in the moderate coupling regime.
Moreover, the strong interaction between the gating qubit and gating thermal bath is unraveled to be compulsory to exhibit these far-from equilibrium features.
\end{abstract}

%\pacs{44.90.+c, 63.22.-m, 05.60.Gg, 05.70.Ln}

%\keywords
% 44.90.+c	other topics in heat transfer
%63.22.-m	Phonons or vibrational states in low-dimensional structures and nanoscale materials
% 05.60.Gg	Quantum transport
% 05.70.Ln	Nonequilibrium and irreversible thermodynamics

\maketitle

\section{Introduction}

Unraveling fundamental mechanisms of nonequilibrium quantum transport through low-dimensional systems has been attracting much attention during the past decades~\cite{sdatta2005,apjauho2007,ydubi2011rmp},
which is proved to be heuristic to render the emergence of molecular electronics~\cite{mratner2013nn,dvirasegal2016ann}, spin caloritronics~\cite{gewbauer2012nm} and quantum biology~\cite{kdorfman2013pnas,mmohseni2014}.
Particularly in phononics~\cite{nbli2012rmp,jren2015aip}, functional operations have been extensively  proposed theoretically, guiding design
of thermal diode, thermal memory, thermal clock and even phonon computers~\cite{bli2004prl,bli2006apl,cwchang2006science,lwang2007prl,lwang2008prl,jswang2008epjb,tchan2014am,jhjiang2015prb,xyshen2015prl}.
Basically, under the temperature bias, energy naturally transfers from a hot source to a cold drain.
By including an additional modulation, proper controlling protocols are exploited to drive heat current against the thermodynamic bias,
or even exhibit heat amplification effect~\cite{nbli2012rmp}.

To control the nonequilibrium heat transfer in quantum devices, there exist two main approaches:
One is to apply a time-dependent driving field to the device or baths/electrodes.
Specifically for the stochastic field~\cite{nasinitsyn2007prl,dsegal2008prl}, the direction of heat flow can be conveniently modulated, and the thermodynamic efficiency may approach the Carnot limit.
For the periodic driving in the adiabatic limit~\cite{jieren2010prl,tyuge2012prb,chentian2013prb,mfl2016prb}, a geometric-phase induced heat pump and quantization of heat current are revealed.
While extended to the nonadiabatic regime~\cite{mstrass2005prl,dsegal2006pre,mrey2007prb,klw2014ptep}, a Floquet theorem is frequently included to analyze the contribution from quasi-modes to heat transfer.
The other one is to add a control gate to establish three-terminal systems, which has been recently intensively investigated in quantum heat engine~\cite{akato2015jcp,dzxu2016njp,gbenenti2017pr}, inelastic thermoelectricity~\cite{jieren2012prb,jhjiang2016crp,jhjiang2017pra} and quantum thermal transistor~\cite{prl2014pbabdallah,kjoulain2016prl,rsanchez2017prb}.
Intriguingly, the negative differential thermal conductance (NDTC) and heat amplification~\cite{bli2006apl}, nonlinear far-from equilibrium features, are exploited.

NDTC is a uncommon behavior, where an increase of the temperature bias between thermal baths results in an abnormal decrease of heat current passing through the quantum device.
It is analogous to negative differential electronic conductance, which was originally pointed out by L. Esaki and his coworkers to study the transport of resonant tunneling diodes in bulk semiconductor devices~\cite{lesaki1958pr,llchang1974apl}.
NDTC has been extensively analyzed in phononic lattices~\cite{nbli2012rmp}, which was revealed to constitute the main ingredient for thermal transistor operation
and devise various logic gates.
Moreover, the NDTC is considered as a crucial component to realize the giant heat amplification,
which describes that the tiny change of base current will significantly modulate the current at collector and emitter.
Recently, a fully quantum thermal transistor, consisting of three coupled qubits is proposed~\cite{kjoulain2016prl}. %, which transfers both the spin-wave and heat flux simultaneously.
In the weak qubit-bath interaction limit, the NDTC and heat amplification are clearly unraveled.
Here, we raise the question: \emph{can we find the NDTC and giant heat amplification by including only energy exchange processes in quantum qubits system?
Considering the crucial effect of qubit-bath interaction in heat transfer in nonequlibrium spin-boson model~\cite{ajleggett1987rmp,dvirasegal2006prb,dzxu2016njp},
what is the influence of strong qubit-bath interaction on these nonequilibrium features?}

In this paper, we investigate steady state heat transfer in a nonequilibrium triangle-coupled spin-boson system, in which only energy is allowed to exchange.
We combine the nonequilibrium noninteracting blip approximation (NIBA) scheme with full counting statistics, detailed in Sec. II, part B,
which is particularly powerful to handle the strong system-bath interaction.
The NDTC and giant heat amplification factor in the three-terminal setup are clearly exploited at strong qubit-bath coupling.
Then, we propose the underlying mechanism to exhibit these far-from equilibrium effects,
and interestingly find that the strong qubit-bath interaction of the gate part is compulsory to generate the NDTC.
The work is organized as follows:
in Sec. II, we introduce the coupled spin-boson model, and apply the nonequilibrium NIBA together with full counting statistics, to obtain the steady state populations and heat currents.
In Sec. III, we investigate the NDTC and heat amplification, and discuss the corresponding mechanisms.
In Sec. IV, we give a brief summary.

%%==========================================
\begin{figure}[tbp]
%\begin{center}
%\vspace{-2.2cm}
\includegraphics[scale=0.35]{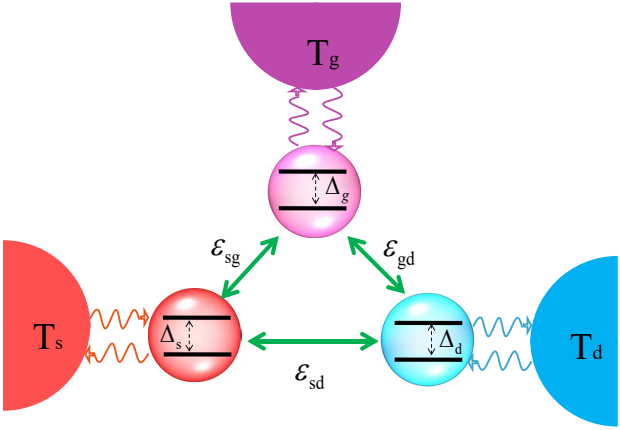}
%\vspace{-2.0cm}
%\end{center}
\caption{(Color online) A schematic description of the nonequilibrium triangle-coupled spin-boson system.
The (red, purple and blue) circles embedded with two dark lines are two-level qubits, $\Delta_v~(v=\textrm{s},\textrm{g},\textrm{d})$
is the intra-qubit tunneling strength between two levels, and $\epsilon_{v,u}~(v{\neq}u)$ is the inter-qubit  interaction
between the $v$th and $u$th qubits.
The (red, purple and blue) half pearls are thermal baths, characterized by temperatures $T_{\textrm{s}}$, $T_{\textrm{g}}$ and $T_{\textrm{d}}$, respectively;
and the arrowed zigzag circles demonstrate interactions between the qubits and thermal baths.
}~\label{fig0a}
\end{figure}
%%==========================================

\section{Model and method}
In this section, we firstly describe triangle-coupled spin-boson system.
Next, we apply the nonequilibrium NIBA scheme to derive the equation of motion of the density matrix of reduced qubits,
and obtain the steady state distribution.
Finally, we show the expression of the steady state energy flux out of the thermal bath.

\subsection{Triangle-coupled spin-boson system}
The nonequilibrium system consisting of three coupled two-level qubits separately interacting with thermal baths, is  described as
$\hat{H}_0=\sum_{v=\textrm{s},\textrm{g},\textrm{d}}\hat{H_v}+\hat{H}_{q}$, which is schematically shown at Fig.~\ref{fig0a}.
The single nonequilibrium spin-boson model (NESB) at the $v$th site, which is a representative paradigm to describe the heat transfer at nanoscale, is expressed as~\cite{ajleggett1987rmp,uweiss2008book}
\begin{eqnarray}~\label{ham0}
\hat{H_v}=\frac{\Delta_v}{2}\hat{\sigma}^v_x
+\sum_{k}[\hat{\sigma}^{v}_z(\lambda_{k,v}\hat{a}^{\dag}_{k}+\lambda^{*}_{k,v}\hat{a}_{k,v})
+\omega_k\hat{a}^{\dag}_{k,v}\hat{a}_{k,v}].
\end{eqnarray}
The $v$th pseudo-Pauli operator is characterized by $\hat{\sigma}^v_{x}=|\uparrow{\rangle}_v{\langle}\downarrow|+|\downarrow{\rangle}_v{\langle}\uparrow|$
and $\hat{\sigma}^v_{z}=|\uparrow{\rangle}_v{\langle}\uparrow|-|\downarrow{\rangle}_v{\langle}\downarrow|$,
and $\Delta_v$ is the tunneling strength of the $v$th qubit.
$\hat{a}^{\dag}_{k,v}~(\hat{a}_{k,v})$ is the creating (annihilating) operator of one boson (e.g., phonon, photon) with frequency $\omega_k$ in the momentum $k$,
and $\lambda_{k,v}$ is the interacting strength between the $v$th qubit and the corresponding thermal bath.
The inter-qubit coupling is given by
\begin{eqnarray}~\label{ham1}
\hat{H}_{q}=\sum_{v{\neq}u}\frac{\epsilon_{vu}}{2}\hat{\sigma}^v_z\hat{\sigma}^u_z,
\end{eqnarray}
where $\epsilon_{vu}$ is coupling strength between the $v$th and $u$th qubits, and $\epsilon_{v,v}=0$.

The spin-boson model was originally introduced to investigate the dissipation of a two-level qubit~\cite{ajleggett1987rmp},
which has been extensively analyzed in quantum decoherence, quantum information measurement and quantum phase transition.
It was later extended to the NESB~\cite{dvirasegal2005prl,dsegal2008prl,ksaito2013prl}.
The qubit-bath inteaction was unraveled to exhibit novel transfer behaviors far-from equilibrium.
Particularly, in the weak qubit-bath coupling limit, the seminal Redfield scheme is adopted to study the
resonant energy transfer of the qubit~\cite{dvirasegal2005prl,jieren2010prl}.
While in the strong coupling regime, the nonequilibrium noninteracting blip approximation (NIBA) is widely considered to analyze the
multi-boson involved scattering processes~\cite{hdekker1987pra,dvirasegal2006prb,tianchen2013prb},
and was recently improved by nonequilibrium polaron-transformed Redfield equation~\cite{cwang2015sr,dzxu2016fp,cwang2017pra}
and Green's function approaches~\cite{jjliu2016cp,jjliu2017pre}.
As is known, the nonequilibrium NIBA will breakdown in the weak qubit-bath coupling limit, compared to the Redfield scheme~\cite{tianchen2013prb,cwang2015sr}.
Hence, in the following, we apply the nonequilibrium NIBA to focus on energy transport of three coupled spin-boson model beyond weak coupling regime.

We include the Silbey-Harris transformation $\hat{U}=\exp{(\frac{i}{2}\sum_{v}\hat{\sigma}^v_z\hat{B}_v)}$~\cite{rsilbey1984jcp,rharrris1985jcp} to obtain the transformed Hamiltonian as $\hat{H}=\hat{U}^{\dag}\hat{H}_0\hat{U}=\hat{H}_q+\hat{H}_b+\hat{V}_{qb}$,
with the collective boson momentum operator at the $v$th bath
$\hat{B}_v=2i\sum_{k}(\frac{\lambda_{k,v}}{\omega_k}\hat{a}^{\dag}_{k,v}-\frac{\lambda^{*}_{k,v}}{\omega_k}\hat{a}_{k,v})$.
After transformation, the transformed system is given by $\hat{H}_q$ at Eq.~(\ref{ham1}),
which is intact due to the commutating relation $[\hat{U},\hat{H}_q]=0$.
The bath is described as
$\hat{H}_b=\sum_{k,v}\omega_k\hat{a}^{\dag}_{k,v}\hat{a}_{k,v}$,
and the qubit-bath interaction is given by
\begin{eqnarray}~\label{vsb1}
\hat{V}_{qb}=\sum_v(e^{-i\hat{B}_v}\hat{\sigma}^v_++e^{i\hat{B}_v}\hat{\sigma}^v_-).
\end{eqnarray}
From
$e^{{\pm}i\hat{B}_v}\hat{\sigma}^v_{\mp}=[1{\pm}i\hat{B}_v-(\hat{B}_v)^2/2+{\cdots}]\hat{\sigma}^v_{\mp}$  at Eq.~(\ref{vsb1}),
high order terms $(\hat{B}_v)^n/n!\hat{\sigma}^v_{\mp}~(n{\ge}2)$ are clear to unravel the multi-boson contribution to the energy transfer, accompanied by the spin flips
$\hat{\sigma}^v_{+}=|\uparrow{\rangle}_v{\langle}\downarrow|$ and $\hat{\sigma}^v_{-}=|\downarrow{\rangle}_v{\langle}\uparrow|$.

\begin{comment}
%%==========================================
\begin{figure}[tbp]
%\begin{center}
%\vspace{-2.2cm}
\includegraphics[scale=0.35]{fig0b.eps}
%\vspace{-2.0cm}
%\end{center}
\caption{(Color online) All possible energy transfer processes based on the kinetic equation at Eq.~(\ref{p1}),
and $\kappa_{ij}$ describes the process from collective population state $\mathcal{P}_j$ to $\mathcal{P}_i$, shown at appendix A.
}~\label{fig0b}
\end{figure}
%%==========================================
\end{comment}

\subsection{Nonequlibrium NIBA}

Here, we apply the nonequilibrium NIBA approach to analyze the steady state energy transfer in the coupled spin-boson system,
which is particularly appropriate in strong qubit-bath coupling  or small intra-qubit tunneling regime~\cite{dvirasegal2006prb,tianchen2013prb}.
By considering the Born-Markov approximation, we expand the nonlinear qubit-bath interaction at Eq.~(\ref{vsb1}) up to the second order,
resulting in the dynamical equation of the reduced qubits density matrix as
\begin{eqnarray}~\label{qme0}
\frac{d}{dt}\hat{\rho}_s&=&-i[\hat{H}_q,\hat{\rho}_s]
+\sum_{v=s,g,d;\eta=\pm}\int^{\infty}_0d{\tau}{\times}\nonumber\\
&&(C_{v}(\tau)[\hat{\sigma}^v_{\eta}(-\tau)\hat{\rho}_s,\hat{\sigma}^v_{\overline{\eta}}]
+H.c.),
\end{eqnarray}
with $\overline{\eta}=-\eta$ and the commutating relation $[\hat{A},\hat{B}]=\hat{A}\hat{B}-\hat{B}\hat{A}$.
The correlation function of the $v$th thermal bath is given by $C_v(\tau)=(\frac{\Delta_v}{2})^2\exp{[-Q_v(\tau)]}$, with the corresponding propagating function
\begin{eqnarray}~\label{q1}
Q_v(\tau)&=&\int^{\infty}_0d{\omega}\frac{J_v(\omega)}{\pi\omega^2}[\coth(\frac{\beta_v\omega}{2})(1-\cos\omega\tau)+i\sin\omega\tau],\nonumber\\
\end{eqnarray}
where spectrum function of the $v$th bath is
$J_v(\omega)=4\pi\sum_k|\lambda_{k,v}|^2\delta(\omega-\omega_k)$,
and the inverse of the temperature is $\beta_v=1/k_bT_v$.
In the present paper, we select the spectral function as the Ohmic case
$J_v(\omega)={\pi\alpha_v\omega}e^{-\omega/\omega_{c,v}}$,
with $\alpha_v$ the coupling strength and $\omega_{c,v}$ the cut-off frequency.
The Ohmic spectrum has been widely considered to mimic the environment, e.g., in quantum dissipation~\cite{ajleggett1987rmp,uweiss2008book}, quantum transport~\cite{ksaito2013prl,akato2015jcp} and quantum phase transition~\cite{pwerner2005prl,lwduan2013jcp,zcai2014prl}.

Next, we introduce the collective populations $\mathcal{P}_1=P_{1}+P_8$, $\mathcal{P}_2=P_{2}+P_7$, $\mathcal{P}_3=P_{3}+P_6$
and $\mathcal{P}_4=P_{4}+P_5$, where the population at the state $|i{\rangle}$ is $P_i={\langle}i|\hat{\rho}_s|i{\rangle}$,
with states specified as
$\{|1{\rangle}=|{\uparrow}{\uparrow}{\uparrow}{\rangle},|2{\rangle}=|{\uparrow}{\uparrow}{\downarrow}{\rangle},
|3{\rangle}=|{\uparrow}{\downarrow}{\uparrow}{\rangle},|4{\rangle}=|{\uparrow}{\downarrow}{\downarrow}{\rangle},
|5{\rangle}=|{\downarrow}{\uparrow}{\uparrow}{\rangle},|6{\rangle}=|{\downarrow}{\uparrow}{\downarrow}{\rangle},
|7{\rangle}=|{\downarrow}{\downarrow}{\uparrow}{\rangle},|8{\rangle}=|{\downarrow}{\downarrow}{\downarrow}{\rangle}\}$.
The eigen-energies corresponding to $\mathcal{P}_i$ are given by
$E_1=\epsilon_{sg}+\epsilon_{sd}+\epsilon_{gd}$,
$E_2=\epsilon_{sg}-\epsilon_{sd}-\epsilon_{gd}$,
$E_3=-\epsilon_{sg}+\epsilon_{sd}-\epsilon_{gd}$ and
$E_4=-\epsilon_{sg}-\epsilon_{sd}+\epsilon_{gd}$.
It should be noted that as we reduce eight-state space ($\{P_i\}$) to four-state case ($\{\mathcal{P}_i\}$),
the spin-reversal symmetry (or say spin-degeneracy) is considered,
which is valid when there is no external field and Zeeman split.
The generalization to the spin-reversal broken case is straightforward, and is not discussed in this paper.

Then, the dynamical equations of collective populations are expressed as
\begin{eqnarray}~\label{p1}
\frac{d\mathcal{P}_i}{dt}&=&-\kappa_{ii}\mathcal{P}_i+\sum_{j{\neq}i}\kappa_{ij}\mathcal{P}_j,
\end{eqnarray}
where the rates $\kappa_{ij}$ describe the transition from $\mathcal{P}_j$ to $\mathcal{P}_i$, which are specified in {\color{blue}appendix A}.
They fulfill the local detail balanced relation between two collective population states $\mathcal{P}_i$ and $\mathcal{P}_j$ as
\begin{eqnarray}
\kappa_{i,j}=\kappa_{j,i}e^{\beta_v(E_j-E_i)},
\end{eqnarray}
in which the process is mediated by $v$th thermal bath.
After long time evolution, the steady state collective populations are obtained as
\begin{comment}
\begin{eqnarray}~\label{pop}
\mathcal{P}^{s}_1&=&(\kappa_{14}\kappa_{22}\kappa_{33}+\kappa_{13}\kappa_{22}\kappa_{34}+\kappa_{12}\kappa_{33}\kappa_{24}\\
&&-\kappa_{14}\kappa_{23}\kappa_{32}+\kappa_{13}\kappa_{32}\kappa_{24}+\kappa_{12}\kappa_{23}\kappa_{34})/|A|,\nonumber
\end{eqnarray}
\begin{eqnarray}
\mathcal{P}^s_{2}&=&(\kappa_{11}\kappa_{33}\kappa_{24}+\kappa_{11}\kappa_{23}\kappa_{34}+\kappa_{14}\kappa_{21}\kappa_{33}\nonumber\\
&&-\kappa_{13}\kappa_{31}\kappa_{24}+\kappa_{14}\kappa_{23}\kappa_{31}+\kappa_{13}\kappa_{21}\kappa_{34})/|A|,\nonumber
\end{eqnarray}
\begin{eqnarray}
\mathcal{P}^s_{3}&=&(\kappa_{11}\kappa_{22}\kappa_{34}+\kappa_{11}\kappa_{32}\kappa_{24}+\kappa_{14}\kappa_{22}\kappa_{31}\nonumber\\
&&-\kappa_{12}\kappa_{21}\kappa_{34}+\kappa_{14}\kappa_{21}\kappa_{32}+\kappa_{12}\kappa_{31}\kappa_{24})/|A|,\nonumber
\end{eqnarray}
\end{comment}
\begin{eqnarray}~\label{pop}
\mathcal{P}^{s}_1&=&[\kappa_{14}(\kappa_{12}+\kappa_{32}+\kappa_{42})(\kappa_{13}+\kappa_{23}+\kappa_{43})\\
&&+\kappa_{13}\kappa_{34}(\kappa_{12}+\kappa_{32}+\kappa_{42})+\kappa_{12}\kappa_{24}(\kappa_{13}+\kappa_{23}+\kappa_{43})\nonumber\\
&&-\kappa_{14}\kappa_{23}\kappa_{32}+\kappa_{13}\kappa_{32}\kappa_{24}+\kappa_{12}\kappa_{23}\kappa_{34}]/|A|,\nonumber
\end{eqnarray}
\begin{eqnarray}
\mathcal{P}^s_{2}&=&[(\kappa_{21}+\kappa_{31}+\kappa_{41})(\kappa_{13}+\kappa_{23}+\kappa_{43})\kappa_{24}\\
&&+(\kappa_{21}+\kappa_{31}+\kappa_{41})\kappa_{23}\kappa_{34}+\kappa_{14}\kappa_{21}(\kappa_{13}+\kappa_{23}+\kappa_{43})\nonumber\\
&&-\kappa_{13}\kappa_{31}\kappa_{24}+\kappa_{14}\kappa_{23}\kappa_{31}+\kappa_{13}\kappa_{21}\kappa_{34}]/|A|,\nonumber
\end{eqnarray}
\begin{eqnarray}
\mathcal{P}^s_{3}&=&[(\kappa_{21}+\kappa_{31}+\kappa_{41})(\kappa_{12}+\kappa_{32}+\kappa_{42})\kappa_{34}\\
&&+(\kappa_{21}+\kappa_{31}+\kappa_{41})\kappa_{32}\kappa_{24}+\kappa_{14}\kappa_{31}(\kappa_{12}+\kappa_{32}+\kappa_{42})\nonumber\\
&&-\kappa_{12}\kappa_{21}\kappa_{34}+\kappa_{14}\kappa_{21}\kappa_{32}+\kappa_{12}\kappa_{31}\kappa_{24}]/|A|,\nonumber
\end{eqnarray}
and $\mathcal{P}^s_4=1-\sum_{i=1,2,3}\mathcal{P}^s_i$, with the coefficient
$|A|=(\kappa_{21}+\kappa_{31}+\kappa_{41}+\kappa_{14})[(\kappa_{24}+\kappa_{12}+\kappa_{32}+\kappa_{42})(\kappa_{34}+\kappa_{13}+\kappa_{23}+\kappa_{43})
-(\kappa_{24}-\kappa_{23})(\kappa_{34}-\kappa_{32})]
-(\kappa_{14}-\kappa_{12})[(\kappa_{24}-\kappa_{21})(\kappa_{34}+\kappa_{13}+\kappa_{23}+\kappa_{43})-(\kappa_{24}-\kappa_{23})(\kappa_{34}-\kappa_{31})]
+(\kappa_{14}-\kappa_{13})[(\kappa_{24}-\kappa_{21})(\kappa_{34}-\kappa_{32})-(\kappa_{24}+\kappa_{12}+\kappa_{32}+\kappa_{42})(\kappa_{34}-\kappa_{31})]$.

\subsection{Steady state energy flux}
Here we combine the nonequilibrium NIBA with full counting statistics~\cite{mesposito2009rmp} (see details at {\color{blue}appendix B}) to obtain the steady state energy flux into of the $v$th bath.
To count the quanta of energy into the $v$th thermal bath, we add the counting filed parameter into $\hat{H}_0$ as
$\hat{H}_0(\{\chi\})=e^{i\sum_v\hat{H}^v_b\chi_v/2}\hat{H}_0e^{-i\sum_v\hat{H}^v_b\chi_v/2}=\sum_{v=s,g,d}\hat{H}_v(\chi_v)+\hat{H}_q$,
with the parameter set $\{\chi\}=(\chi_s,\chi_g,\chi_d)$.
The modified $v$th spin-boson model is expressed as
\begin{eqnarray}~\label{hchi0}
\hat{H_v}(\chi_v)&=&\frac{\Delta_v}{2}\hat{\sigma}^v_x+\sum_{k}\omega_k\hat{a}^{\dag}_{k,v}\hat{a}_{k,v}\\
&&+\sum_{k}\hat{\sigma}^{v}_z(\lambda_{k,v}e^{i\omega_k\chi_{v}/2}\hat{a}^{\dag}_{k}+\lambda^{*}_{k,v}e^{-i\omega_k\chi_v/2}\hat{a}_{k,v}).\nonumber
\end{eqnarray}
Next, we applying a modified Silbey-Harris transformation $\hat{U}(\{\chi\})=\exp[\frac{i}{2}\sum_v\hat{\sigma}^v_z\hat{B}_v({\chi}_v)]$
to $\hat{H}_0(\{\chi\})$
as $\hat{H}(\{\chi\})=\hat{U}^{\dag}(\{\chi\})\hat{H}_0(\{\chi\})\hat{U}(\{\chi\})$,
with the collective boson momentum
$\hat{B}_v({\chi}_v)=2i\sum_k(\frac{\lambda_{k}}{\omega_k}e^{i\omega_k\chi_v/2}\hat{a}^{\dag}_{k,v}
-\frac{\lambda^{*}_{k,v}}{\omega_k}e^{-i\omega_k\chi_v/2}\hat{a}_{k,v})$.
Then, the transformed Hamiltonian combined with the counting field parameters is given by
\begin{eqnarray}
\hat{H}(\{\chi\})=\hat{H}_q+\hat{H}_b+\hat{V}_{qb}(\{\chi\}),
\end{eqnarray}
where the qubit-bath interaction is
\begin{eqnarray}~\label{v1}
\hat{V}_{qb}(\{\chi\})=\sum_v(e^{-i\hat{B}_v(\chi_v)}\hat{\sigma}^v_++e^{i\hat{B}_v(\chi_v)}\hat{\sigma}^v_-),
\end{eqnarray}
which reduces to Eq.~(\ref{vsb1})  in absence of counting field parameters.
Based on the Born-Markov approximation, we obtain the master equation of the qubits system by expanding $\hat{V}_{sb}(\{\chi\})$ up to the second order as
\begin{eqnarray}~\label{qmechi1}
\frac{d}{dt}\hat{\rho}_{\chi}&=&-i[\hat{H}_s,\hat{\rho}_{\chi}]
-\sum_{v,\eta}\int^{\infty}_0d{\tau}[C_v(\tau)\hat{\sigma}^v_{\eta}\hat{\sigma}^v_{\overline{\eta}}(-\tau)\hat{\rho}_{\chi}+H.c.]\nonumber\\
&&+\sum_{v,\eta}[C_{\chi_v}(\tau)\hat{\sigma}^v_{\eta}(-\tau)\hat{\rho}_{\chi}\hat{\sigma}^v_{\overline{\eta}}
+C_{\chi_v}(-\tau)\hat{\sigma}^v_{\eta}\hat{\rho}_{\chi}\hat{\sigma}^v_{\overline{\eta}}(-\tau)]\nonumber\\
\end{eqnarray}
where the correlation function is $C_{\chi_v}(\tau)=(\frac{\Delta_v}{2})^2\exp[-Q_{\chi_v}(\tau)]$, and the boson propagator is
\begin{eqnarray}~\label{qchi1}
Q_{\chi_v}(\tau)&=&\int^{\infty}_0d{\omega}\frac{J_v(\omega)}{\pi\omega^2}\{\coth(\frac{\beta_v\omega}{2})(1-\cos[\omega(\tau-\chi_v)])\nonumber\\
&&+i\sin[\omega(\tau-\chi_v)]\}.
\end{eqnarray}
In absence of the counting parameter $\chi_v=0$,
the quantum master equation at Eq.~(\ref{qmechi1}) returns back to the standard version at Eq.~(\ref{qme0}), and the propagator at Eq.~(\ref{qchi1}) reduces to Eq.~(\ref{q1}).

Then, we define the collective populations as
$\mathcal{P}_1(\{\chi\})=P_1(\{\chi\})+P_8(\{\chi\})$,
$\mathcal{P}_2(\{\chi\})=P_2(\{\chi\})+P_7(\{\chi\})$,
$\mathcal{P}_3(\{\chi\})=P_3(\{\chi\})+P_6(\{\chi\})$,
and $\mathcal{P}_4(\{\chi\})=P_4(\{\chi\})+P_5(\{\chi\})$.
The dynamical equation can be expressed as
\begin{eqnarray}~\label{qme1}
\frac{d}{dt}\mathcal{P}_i(\{\chi\})=-\kappa_{ii}\mathcal{P}_i(\{\chi\})+\sum_{j{\neq}i}\kappa_{ij}(\{\chi\})\mathcal{P}_j(\{\chi\}),
\end{eqnarray}
where the modified transition rates are specified in {\color{blue} appendix A}. In absence of counting field parameter set $\{\chi=0\}$, it becomes equivalent with Eq.~(\ref{p1}).
It should be noted that as $\chi_v{\neq}0$, the detailed balanced relation breaks down.
If we arrange the population vector as $|P(\{\chi\}){\rangle}=[\mathcal{P}_1(\{\chi\}),\mathcal{P}_2(\{\chi\}),\mathcal{P}_3(\{\chi\}),\mathcal{P}_4(\{\chi\})]$,
the dynamical equation is re-expressed as
$\frac{d}{dt}|P(\{\chi\}){\rangle}=-\mathcal{\hat{L}}(\{\chi\})|P(\{\chi\}){\rangle}$,
with $\mathcal{\hat{L}}(\{\chi\})$ the superoperator built from the modified transition rates $\kappa_{ij}(\{\chi\})$.

Finally, the steady state energy current out of the $v$th thermal bath is given by
\begin{eqnarray}
J_v={\langle}I|\frac{{\partial}\mathcal{\hat{L}}(\{\chi\})}{{\partial}(i\chi_v)}|_{\{\chi=0\}}|\mathcal{P}_{s}{\rangle},
\end{eqnarray}
where the unit vector ${\langle}I|=[1,1,1,1]$, and $|\mathcal{P}_{s}{\rangle}$ is the steady state.
Specifically, the steady state currents out of the source thermal bath is given by
\begin{eqnarray}~\label{jl1}
J_s&=&2[(\epsilon_{sg}+\epsilon_{sd})(\kappa_{14}\mathcal{P}^{s}_4-\kappa_{41}\mathcal{P}^{s}_1)\nonumber\\
&&+(\epsilon_{sg}-\epsilon_{sd})(\kappa_{23}\mathcal{P}^{s}_3-\kappa_{32}\mathcal{P}^{s}_2)].
\end{eqnarray}
$J_s$ originates from the energy exchange between the source bath and the corresponding qubit.
%From Fig.~\ref{fig0b}, it is known that two transitions completely contribute to this current: $\mathcal{P}^s_1{\rightleftharpoons}\mathcal{P}^s_4$ and $\mathcal{P}^s_2{\rightleftharpoons}\mathcal{P}^s_3$, with the energy exchange for each transition $E_1-E_4$ and $E_2-E_3$, respectively.
It should be noted that thermal baths individually contribute to the current,
and heat transfer between two baths should be mediated by inter-qubit interaction at Eq.~(\ref{ham1}).
Hence, the transition processes are quite different from the single qubit under nonequilibrium condition~\cite{dvirasegal2006prb,tianchen2013prb},
in which the joint contribution of thermal baths occurs to the current.
Similarly, the current out of the gate and source baths are expressed as
\begin{eqnarray}~\label{jm1}
J_g&=&2[(\epsilon_{sg}+\epsilon_{gd})(\kappa_{13}\mathcal{P}^{s}_3-\kappa_{31}\mathcal{P}^{s}_1)\nonumber\\
&&+(\epsilon_{sg}-\epsilon_{gd})(\kappa_{24}\mathcal{P}^{s}_4-\kappa_{42}\mathcal{P}^{s}_2)],
\end{eqnarray}
and
\begin{eqnarray}~\label{jr1}
J_d&=&2[(\epsilon_{sd}+\epsilon_{gd})(\kappa_{12}\mathcal{P}^{s}_2-\kappa_{21}\mathcal{P}^{s}_1)\nonumber\\
&&+(\epsilon_{sd}-\epsilon_{gd})(\kappa_{34}\mathcal{P}^{s}_4-\kappa_{43}\mathcal{P}^{s}_3)],
\end{eqnarray}
respectively.
They fulfill the current reservation $J_s+J_g+J_d=0$.

\section{Results and discussions}
In this section, we first show the NDTC and heat amplification at strong qubit-bath coupling.
Then, we describe the underlying mechanism to exhibit such far-from equilibrium features.

\subsection{Far-from equilibrium effects}

%%==========================================
\begin{figure}[tbp]
%\begin{center}
%\vspace{-2.2cm}
\includegraphics[scale=0.5]{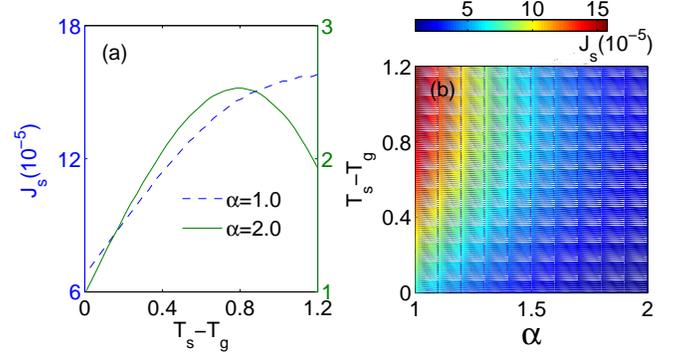}
%\vspace{-2.0cm}
%\end{center}
\caption{(Color online) (a) Energy current out of the $s$th bath $J_s$ by modulating the temperature $T_g$ of the gating bath
at moderate ($\alpha=1.0$) and strong ($\alpha=2.0$) qubit-bath coupling regimes, with $\alpha_s=\alpha_g=\alpha_d=\alpha$;
(b) a birdview of $J_s$ by tuning $\alpha$ and $T_g$.
The other system parameters are given by $\Delta=0.1$, $\epsilon_{sg}=\epsilon_{gd}=1$, $\epsilon_{sd}=0$, $T_s=1.6$, $T_d=0.4$ and $\omega_c=10$.
}~\label{fig1}
\end{figure}

%%==========================================
\subsubsection{Negative differential thermal conductance}

In a two-terminal setup, the phenomenon that current becomes suppressed by increasing the thermodynamic (e.g., voltage and temperature) bias between two baths, is traditionally characterized as negative differential conductance~\cite{oaluf2012}, which has been widely applied to study the nonequilibrium electron transport~\cite{mgalperin2005nl},
and later introduced in phononic functional systems with the similar concept (NDTC)~\cite{dvirasegal2006prb,bli2006apl}.
Recently, this concept is intensively extended to the gate-controlled three-terminal devices~\cite{nbli2012rmp,acsnano2012yqwu,prl2014pbabdallah,kjoulain2016prl}.

Here, we investigate the gate-controlled NDTC feature of the source current $J_s$ by tuning the qubit-bath interaction, shown at Fig.~\ref{fig1}(a).
In the moderate qubit-bath coupling limit (e.g., $\alpha=1.0$ with dashed blue line), the heat current is expectedly found to increase monotonically by
increasing the thermodynamic bias ($T_s-T_g$) between the source and the gate.
However, in the strong qubit-bath interaction regime (e.g., $\alpha=2.0$ with solid green line),
when we tune the temperature bias at $T_s-T_g{\lesssim}0.8$, the heat current exhibits the enhancement.
As the temperature bias further increases, the heat current is astonishingly suppressed.
This clearly demonstrates the existence of the NDTC, which is in sharp contrast to the counterpart in moderate coupling case.
This is one central point in this paper.
To give a comprehensive picture of the NDTC at strong coupling regime, we exhibit a birdview of $J_s$ at Fig.~\ref{fig1}(b). It is interesting to find that the NDTC emerges as the qubit-bath interaction strength increases to $\alpha{\approx}1.5$.
%Hence, we conclude that the NDTC behavior is completely unraveled.

It is necessary to note that in the previous investigation of NESB~\cite{dvirasegal2006prb} with the two terminal setup, the NDTC was exploited at strong qubit-bath interaction regime
by including the NIBA combined the Marcus approximation.
However, the existence of the NDTC was later clarified to be a fake by using the nonequilibrium polaron transformed Redfield scheme~\cite{cwang2015sr},
mainly due to breakdown of the Marcus treatment in the low temperature regime.
Here, with the three terminal setup at Fig.~\ref{fig0a}, the NDTC is definitely true within the framework of NIBA,
in which the transfer processes are significantly different from the counterpart at Ref.~\cite{dvirasegal2006prb}.

%%==========================================
\begin{figure}[tbp]
%\begin{center}
%\vspace{-2.2cm}
\includegraphics[scale=0.45]{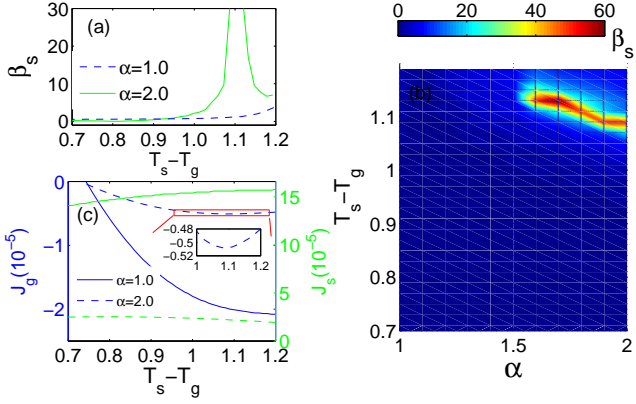}
%\vspace{-2.0cm}
%\end{center}
\caption{(Color online) (a) Heat amplification factor $\beta_s$ at Eq.~(\ref{factor});
(b) a birdview of $\beta_s$ by modulating $T_g$ and  $\alpha$;
and (c) heat current $J_g$ and $J_s$ by varying the temperature $T_g$ in the moderate ($\alpha=1.0$ with blue and green solid lines)
and strong ($\alpha=2.0$ with blue and green dashed lines) coupling regimes respectively, with $\alpha_v=\alpha~(v=s,g,d)$.
The inset in Figure (c) is the zoom-in view of $J_g$ with $\alpha=2.0$.
The other parameters are given by $\Delta=0.1$, $\epsilon_{sg}=\epsilon_{gd}=1$, $\epsilon_{sd}=0$, $T_s=1.6$, $T_d=0.4$ and $\omega_c=10$.
}~\label{fig2}
\end{figure}
%%==========================================

\subsubsection{Heat amplification}

The ability to amplify heat flow constitutes one important ingredient for the operation of three-terminal devices,
particular quantum thermal transistor~\cite{nbli2012rmp,prl2014pbabdallah,kjoulain2016prl}.
The heat amplification factor can be described by the change of the heat current $J_s$ (or $J_d$) upon
the change of gate current $J_g$, which plays the role of control.
It can be explicitly expressed as
\begin{eqnarray}~\label{factor}
\beta_v=|{\partial}J_{v}/{\partial}J_g|,~v=s,d.
\end{eqnarray}
Due to the current conservation $\sum_{v=s,g,d}J_v=0$, the expression of $\beta_d$ can be alternatively expressed as
\begin{eqnarray}~\label{betad}
\beta_d=|\beta_s+(-1)^{\theta}|,
\end{eqnarray}
with $\theta=0$ for $\partial J_s/\partial J_g>0$, and $\theta=1$ for $\partial J_s/\partial J_g<0$.
Generally, when $\beta_{s,d}>1$, we say the amplification effect works~\cite{nbli2012rmp}. Therefore if $\beta_s>2$, we will always have $\beta_d>1$ and both amplification factors larger than 1.
%{\color{red} This expression and the explanation is not simple. Let's make it as simple as possible.}

We emphasize that the NDTC is compulsory for realizing the heat amplification. This is because alternatively
\begin{equation}~\label{betad2}
\beta_d=\left|\frac{\partial J_d}{\partial(J_d+J_s)}\right|=\left|\frac{G_d}{G_d+G_s}\right|,
\end{equation}
where $G_{d,s}=\partial J_{d,s}/\partial T_g$ are the differential thermal conductance at drain and source terminals, respectively. Clearly, to achieve amplification factor $\beta_d$ larger than 1, we need just either $G_d$ negative or $G_s$ negative, not both.

We study the behavior of heat amplification factor $\beta_s$ at Fig.~\ref{fig2}(a).
In the moderate qubit-bath coupling regime (e.g., $\alpha=1.0$ with dashed blue line),
the low heat amplification factor ($\beta_s{\lesssim}4$) is observed in the large temperature bias limit ($T_s-T_g{\approx}1.2$).
While in the strong coupling regime (e.g., $\alpha=2.0$ with solid green line), it is interesting to find that the heat amplification factor is significantly enhanced by increasing temperature bias, and then becomes divergent around $T_s-T_g{\approx}1.1$.
$\beta_d$ from Eq.~(\ref{betad}) also shows divergent behavior, though not plotted here.
Hence, we conclude that strong qubit-bath interaction is crucial to exhibit the large heat amplification factor.
This is the other central point in this paper.
Moreover, we exhibit a comprehensive picture of $\beta_s$ at Fig.~\ref{fig2}(b).
Giant heat amplification factor is clearly shown in strong qubit-bath coupling regime (e.g., $\alpha{\ge}1.5$),
which is consistent with the result at Fig.~\ref{fig2}(a).
%Interestingly, under large temperature bias ($T_s-T_g{\ge}1.0$), the amplification effect is, however, dramatically suppressed at moderate qubit-bath coupling regime,
%which is distinct from the current robustness at Fig.~\ref{fig1}(b).

We also briefly analyze the origin of the amplification factor divergence.
Through analysis of $J_g$ at Fig.~\ref{fig2}(c), it is found that $J_g$ at strong qubit-bath coupling exhibits the NDTC feature around $T_s-T_g{\approx}1.1$,
which is clearly demonstrated at the inset.
Near this critical temperature, the change of $J_g$ is almost negligible (${\partial}J_g/{\partial}T_g{\approx}0$).
Meanwhile, $J_s$ shows monotonic decrease at this critical temperature regime, which is consistent with the finite change of $J_s$ at Fig.~\ref{fig1}(a).
Hence, it results in the novelly divergent behavior of the current amplification factor both for $\beta_s$ and $\beta_d$.
While for $\beta_s$ at the moderate coupling case ($\alpha=1.0$),
though NDTC of $J_g$ disappears, the turnover behavior of $J_d$ can indeed be found in the large bias limit of $T_s-T_g$ (not shown here), which
results in $G_s{\times}G_d<0$.
Thus, according to the Eq.~(\ref{betad2}), $\beta_d$ is able to exceed $2$, which finally makes $\beta_s>1$.

\subsection{Mechanism of the NDTC}

%%==========================================
\begin{figure}[tbp]
%\begin{center}
%\vspace{-2.2cm}
\includegraphics[scale=0.5]{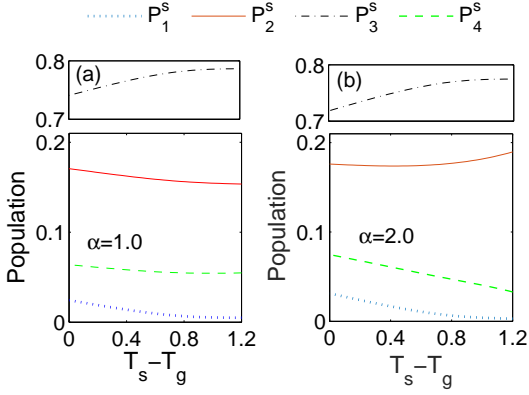}
%\vspace{-2.0cm}
%\end{center}
\caption{(Color online) Steady state collective populations $\mathcal{P}^s_i$ at (a) moderate and (b) strong coupling regimes.
The other parameters are given by $\Delta=0.1$, $\epsilon_{sg}=\epsilon_{gd}=1$, $\epsilon_{sd}=0$, $T_s=1.6$, $T_d=0.4$ and $\omega_c=10$.
}~\label{fig3}
\end{figure}
%%==========================================

\subsubsection{Anomalous behavior of collective populations}

We include the expression of $J_s$ at Eq.~(\ref{jl1}) to exploit the underlying mechanism of the NDTC behavior shown as Fig.~\ref{fig1}.
Under the condition $\epsilon_{sd}=0$ and $\epsilon_{sg}=\epsilon_{gd}=\epsilon$, $J_l$ is specified as
$J_s=2\epsilon(\kappa_{23}\mathcal{P}_3+\kappa_{14}\mathcal{P}_4-\kappa_{41}\mathcal{P}_1+\kappa_{32}\mathcal{P}_2)$.
Moreover, the transition rates $\kappa_{23(32)}$ and $\kappa_{14(41)}$ describe transfer processes mediated by  the $s$th bath,
which will not change with tuning $T_g$.
For  the moderate qubit-bath coupling, the occupation probability $\mathcal{P}_1$ at Fig.~\ref{fig3}(a) becomes negligible,
which results in $J_s{\approx}2\epsilon(\kappa_{23}\mathcal{P}_3-\kappa_{32}\mathcal{P}_2+\kappa_{14}\mathcal{P}_4)$.
By increasing bias $T_s-T_g$, it is clear to see that $\mathcal{P}_3$ shows monotonic increase, whereas $\mathcal{P}_2$ is suppressed.
Moreover, $\mathcal{P}_4$ is nearly unchanged. Hence, these behaviors collectively enhance $J_s$.
However, in the strong coupling regime, $\mathcal{P}_2$ and $\mathcal{P}_3$ dominate $J_l$.
The corresponding expression of $J_s$ can  be further simplified as
\begin{eqnarray}~\label{jls1}
{J}_s{\approx}2\epsilon(\kappa_{23}\mathcal{P}_3-\kappa_{32}\mathcal{P}_2).
\end{eqnarray}
From Fig.~\ref{fig1}(b), it is interesting to found that in the high temperature regime of $T_g$ (e.g., $T_g{\gtrsim}0.8$ and $T_s-T_g\in (0, 0.8)$ for $T_s=1.6$),
$\mathcal{P}_3$ increases dramatically, whereas $\mathcal{P}_2$ keeps nearly stable.
This process will absolutely enhance $J_s$.
While in the comparatively low temperature regime of $T_g$ (e.g., $T_g<0.4$ and $T_s-T_g\in (0.8, 1.2)$ for $T_s=1.6$), $\mathcal{P}_2$ begins to arise,
whereas $\mathcal{P}_3$ reaches a stable value.
Thus, $J_s$ exhibits monotonic decrease, accordingly.
Hence, we conclude that the strong qubit-bath coupling is crucial to exhibit NDTC behavior of $J_s$.
Meanwhile, one question naturally arises: Should all qubit-bath interactions be strong? Or we only keep the gating qubit-bath coupling strong?

%%==========================================
\begin{figure}[tbp]
%\begin{center}
%\vspace{-2.2cm}
\includegraphics[scale=0.45]{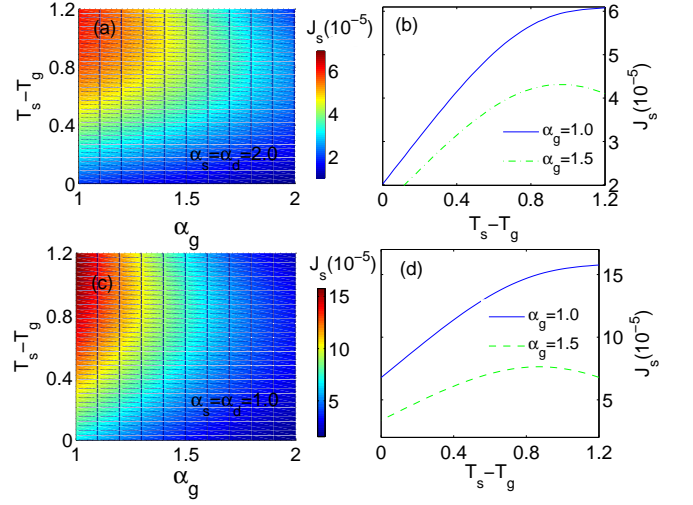}
%\vspace{-2.0cm}
%\end{center}
\caption{(Color online) The behavior of heat current $J_s$ by tuning the gating qubit-bath coupling strength $\alpha_g$ and temperature $T_g$,
with (a),(b) strong source/drain qubit-bath interaction strength $\alpha_s=\alpha_d=2.0$,
and (c), (d) moderate source/drain qubit-bath interaction strength $\alpha_s=\alpha_d=1.0$.
The other parameters are given by $\Delta=0.1$, $\epsilon_{sg}=\epsilon_{gd}=1$, $\epsilon_{sd}=0$, $T_s=1.6$, $T_d=0.4$ and $\omega_c=10$.
}~\label{fig4}
\end{figure}
%%==========================================

\subsubsection{Importance of gating qubit-bath interaction}

We focus on the heat current $J_s$  by modulating the qubit-bath coupling strengthes separately.
We firstly tune coupling strengthes $\alpha_s$ and $\alpha_d$ sufficiently strong to expect to exhibit the NDTC feature, shown at Fig.~\ref{fig4}(a).
It is disappointed to see that for the moderate $\alpha_g$, no signal of the NDTC emerges. Only $\alpha_g{\gtrsim}1.2$, such novel(turnover) behavior begins to occur.
To see it clearly, we plot $J_s$ with typical coupling strengths $\alpha_g$ at Fig.~\ref{fig4}(b). For  the mediate coupling case, $J_s$ shows monotonic increase.
While in the strong coupling regime, the turnover feature is observed.
Hence, we propose that strong qubit-bath interactions for source and drain baths are not necessary.
Next, we release the source and drain bath-qubit couplings to the moderate case ($e.g., \alpha_s=\alpha_d=1.0$), shown at Fig.~\ref{fig4}(c).
It is demonstrated that NDTC behavior also exists for $\alpha_g{\gtrsim}1.2$, which is also confirmed by Fig.~\ref{fig4}(d).
Therefore, we conclude that the strong interaction between the gating qubit and corresponding bath is crucial to exhibit the NDTC behavior, so as to the heat amplification.

Recently, the NDTC was also discovered in a seemingly similar model~\cite{kjoulain2016prl} with weak qubit-bath interactions,
which instead describes the qubit-bath coupling as $\hat{V}_{sb}=\sum_{k,v}\hat{\sigma}^v_{x}(g_{k,v}\hat{a}^{\dag}_{k,v}+g^{*}_{k,v}\hat{a}_{k,v})$.
It exchanges the energy and particle (spin excitation/relaxation) simultaneously,
which is apparently different from the counterpart in this paper with only energy exchange at Eq.~(\ref{ham0}).
Moreover, we consider the intra-qubit tunneling, whereas the Zeeman splitting is included in Ref.~\cite{kjoulain2016prl}.
Hence, the physical processes are significantly distinct from each other.

\section{Conclusion}
In a brief summary, we have investigated the steady state heat transfer in a nonequilibrium triangle-coupled spin-boson system with only energy-exchange.
The nonequilibrium NIBA combined with full counting statistics has been applied to non-weakly perturb the qubit-bath interaction,
and the expressions of steady state populations at Eq.~(\ref{pop}) and heat currents at Eqs.~(\ref{jl1}-\ref{jr1}) have been analytically obtained.
In particular, we have clearly observed the negative differential thermal conductance of $J_s$ with strong qubit-bath coupling,
shown at Fig.~\ref{fig1}.
It is mainly attributed to the anomalous response of the occupation probability $\mathcal{P}_2$ to the temperature $T_g$ of the middle bath.
Moreover, a giant heat amplification factor has been discovered at Fig.~\ref{fig2} with strong qubit-bath interaction at large temperature bias,
which originates from the NDTC feature of $J_g$.
Finally, we have identified that the strong gating qubit-bath coupling is crucial to exhibit the negative differential thermal conductance, so as to the heat amplification.
We believe these findings may provide physical insight for the design of  quantum thermal logic devices.
Moreover, the exploitation of the possible NDTC and heat amplification in the weak coupling limit may be conducted in future, as treated in the Ref.~\cite{kjoulain2016prl} by applying the Redfield scheme.

\section{Acknowledgement}

Chen Wang is supported by the National Natural Science Foundation of China under Grant Nos. 11704093 and 11547124.
Jie Ren acknowledges the National Youth 1000 Talents Program in China, the startup grant at Tongji University and National Natural Science Foundation of China under Grant No. 11775159.

\appendix
\section{Transition rates at Eq.~(\ref{p1}) and Eq.~(\ref{qme1})}

At Eq.~(\ref{p1}), the transition rates $\kappa_{ij}$ are specified as
\begin{eqnarray}
\kappa_{12}&=&\int^{\infty}_{-\infty}d{\tau}[C_d(-\tau)e^{2i(\epsilon_{sd}+\epsilon_{gd})\tau}],\nonumber\\
\kappa_{13}&=&\int^{\infty}_{-\infty}d{\tau}[C_g(-\tau)e^{2i(\epsilon_{sg}+\epsilon_{gd})\tau}],\nonumber\\
\kappa_{14}&=&\int^{\infty}_{-\infty}d{\tau}[C_s(-\tau)e^{2i(\epsilon_{sg}+\epsilon_{sd})\tau}],\nonumber
\end{eqnarray}
\begin{eqnarray}
\kappa_{21}&=&\int^{\infty}_{-\infty}d{\tau}[C_d(-\tau)e^{-2i(\epsilon_{sd}+\epsilon_{gd})\tau}],\nonumber\\
\kappa_{23}&=&\int^{\infty}_{-\infty}d{\tau}[C_s(-\tau)e^{2i(\epsilon_{sg}-\epsilon_{sd})\tau}],\nonumber\\
\kappa_{24}&=&\int^{\infty}_{-\infty}d{\tau}[C_g(-\tau)e^{2i(\epsilon_{sg}-\epsilon_{gd})\tau}],\nonumber
\end{eqnarray}
\begin{eqnarray}
\kappa_{31}&=&\int^{\infty}_{-\infty}d{\tau}[C_g(-\tau)e^{-2i(\epsilon_{sg}+\epsilon_{gd})\tau}],\nonumber\\
\kappa_{32}&=&\int^{\infty}_{-\infty}d{\tau}[C_s(-\tau)e^{-2i(\epsilon_{sg}-\epsilon_{sd})\tau}],\nonumber\\
\kappa_{34}&=&\int^{\infty}_{-\infty}d{\tau}[C_d(-\tau)e^{2i(\epsilon_{sd}-\epsilon_{gd})\tau}],\nonumber
\end{eqnarray}
\begin{eqnarray}
\kappa_{41}&=&\int^{\infty}_{-\infty}d{\tau}[C_s(-\tau)e^{-2i(\epsilon_{sg}+\epsilon_{sd})\tau}],\nonumber\\
\kappa_{42}&=&\int^{\infty}_{-\infty}d{\tau}[C_g(-\tau)e^{-2i(\epsilon_{sg}-\epsilon_{gd})\tau}],\nonumber\\
\kappa_{43}&=&\int^{\infty}_{-\infty}d{\tau}[C_d(-\tau)e^{-2i(\epsilon_{sd}-\epsilon_{gd})\tau}],\nonumber
\end{eqnarray}
$\kappa_{11}=\kappa_{21}+\kappa_{31}+\kappa_{41}$,
$\kappa_{22}=\kappa_{12}+\kappa_{32}+\kappa_{42}$,
$\kappa_{33}=\kappa_{13}+\kappa_{23}+\kappa_{43}$,
and
$\kappa_{44}=\kappa_{14}+\kappa_{24}+\kappa_{34}$.
While at Eq.~(\ref{qme1}), the modified transition rates with counting field parameters are specified as
$\kappa_{12}(\chi_d)=e^{-2i(\epsilon_{sd}+\epsilon_{gd})\chi_d}\kappa_{12}$,
$\kappa_{13}(\chi_g)=e^{-2i(\epsilon_{sg}+\epsilon_{gd})\chi_g}\kappa_{13}$ and
$\kappa_{14}(\chi_s)=e^{-2i(\epsilon_{sg}+\epsilon_{sd})\chi_s}\kappa_{14}$;
$\kappa_{21}(\chi_d)=e^{2i(\epsilon_{sd}+\epsilon_{gd})\chi_d}\kappa_{21}$,
$\kappa_{23}(\chi_s)=e^{-2i(\epsilon_{sg}-\epsilon_{sd})\chi_s}\kappa_{23}$ and
$\kappa_{24}(\chi_g)=e^{-2i(\epsilon_{sg}-\epsilon_{gd})\chi_g}\kappa_{24}$;
$\kappa_{31}(\chi_g)=e^{2i(\epsilon_{sg}+\epsilon_{gd})\chi_g}\kappa_{31}$,
$\kappa_{32}(\chi_s)=e^{2i(\epsilon_{sg}-\epsilon_{sd})\chi_s}\kappa_{32}$ and
$\kappa_{34}(\chi_d)=e^{-2i(\epsilon_{sd}-\epsilon_{gd})\chi_d}\kappa_{34}$;
$\kappa_{41}(\chi_s)=e^{2i(\epsilon_{sg}+\epsilon_{sd})\chi_s}\kappa_{41}$,
$\kappa_{42}(\chi_g)=e^{2i(\epsilon_{sg}-\epsilon_{gd})\chi_d}\kappa_{42}$ and
$\kappa_{43}(\chi_d)=e^{2i(\epsilon_{sd}-\epsilon_{gd})\chi_g}\kappa_{43}$.
In absence of the counting parameters ($\chi_v=0$), these transition rates reduce to the standard rates given at Appendix A, accordingly.

\section{brief introduction of full counting statistics}
Considering the heat transfer from the qubits system to the $v$th thermal bath during a finite time $\tau$,
the quanta of transferred heat $\Delta{q}_{\tau}$ is expressed as ${\Delta}q^v_{\tau}=\sum_k\omega_k\Delta{n_{k,v}}$,
where $\omega_k$ is the frequency of phonon in the momentum $k$
and $\Delta{n}_{k,v}=n_{k,v}(\tau)-n_{k,v}(0)$ is the increment of phonon number at $\tau$ time $n_{k,v}(\tau)$
to the initial one $n_{k,v}(0)$.
Specifically, we introduce a projector $\hat{K}_{q^v_{0}}=|q^v_0{\rangle}{\langle}q^v_0|$ to measure the initial quantity
$\hat{H}_v=\sum_{k}\omega_k\hat{b}^{\dag}_{k,v}\hat{b}_{k,v}$, with $q^v_0=\sum_k\omega_kn_{k,v}(0)$.
Then at time $\tau$, we again detect $\hat{H}_v$ to obtain $q^v_{\tau}=\sum_k\omega_kn_{k,v}(\tau)$ by using
the projector $\hat{K}_{q^v_{\tau}}=|q^v_{\tau}{\rangle}{\langle}q^v_{\tau}|$, with $q^v_{\tau}=\sum_k\omega_kn_{k,v}(\tau)$.
Hence, the joint probability of this two-time measurement is obtained as~\cite{mesposito2009rmp}
\begin{eqnarray}
\textrm{Pr}(q^v_{\tau},q^v_0)=\textrm{Tr}_{s,b}\{\hat{K}_{q^v_{\tau}}e^{-i\hat{H}_0\tau}\hat{K}_{q^v_0}\hat{\rho}_0\hat{K}^v_{q_0}
e^{i\hat{H}_0\tau}\hat{K}^v_{q_{\tau}}\},
\end{eqnarray}
with $\hat{\rho}_0$ the initial density matrix, and the trace over both the qubits and baths.
%connect cumulant generating function with the probability during time interval tau
Based on the joint probability $\textrm{Pr}(q^v_{\tau},q^v_0)$, we define the probability of the measurement of $\Delta{Q}^v_{\tau}$ during the time interval $\tau$ as
\begin{eqnarray}
\textrm{Pr}_{\tau}(\Delta{Q}^v_{\tau})=\sum_{q^v_{\tau},q^v_0}\delta[\Delta{Q}^v_{\tau}-(q^v_{\tau}-q^v_0)]\textrm{Pr}(q^v_{\tau},q^v_0).
\end{eqnarray}
Then, the cumulant generating function of the current statistics can be defined as
\begin{eqnarray}
G_{\tau}(\chi_v)=\ln{\int{d\Delta{Q}^v_{\tau}}\textrm{Pr}_{\tau}(\Delta{Q}^v_{\tau})e^{i\chi_v\Delta{Q}^v_{\tau}}},
\end{eqnarray}
with $\chi_v$ the counting field parameter relating with the $v$th thermal bath.
Consequently, the heat current is obtained as the first order cumulant case
\begin{eqnarray}
J_v=\frac{{\partial}G(\chi_v)}{{\partial}(i\chi_v)}|_{\chi_v=0},
\end{eqnarray}
with the steady state cumulant generating function $G(\chi_v)=\lim_{\tau{\rightarrow}\infty}G_{\tau}(\chi_v)$.

If the the dynamical equation of the qubits with the counting field parameter in the Liouvillian framework is given by
\begin{eqnarray}
\frac{d}{dt}|P(\chi_v){\rangle}=-\mathcal{\hat{L}}(\chi_v)|P(\chi_v){\rangle}.
\end{eqnarray}
After the long time evolution, the cumulant generating function is simplified as
$G(\chi_v)=E_0(\chi_v)$, with $E_0(\chi_v)$ the eigenvalue of $\mathcal{\hat{L}}(\chi_v)$ owning the maximal real part.
Hence, the heat current can be obtained as $J_v=\frac{{\partial}E_0(\chi_v)}{{\partial}(i\chi_v)}|_{\chi_v=0}$.
Alternatively, the current can also be expressed as
\begin{eqnarray}
J_v={\langle}I|\frac{{\partial}\mathcal{\hat{L}}(\chi_v)}{{\partial}(i\chi_v)}|_{\chi_v=0}|P_{s}{\rangle},
\end{eqnarray}
with ${\langle}I|$ the unit vector and $|P_{ss}{\rangle}$ the steady state of the qubits system in absence of the counting field parameter.

%%==========================================
\begin{figure}[tbp]
%\begin{center}
%\vspace{-2.2cm}
\includegraphics[scale=0.5]{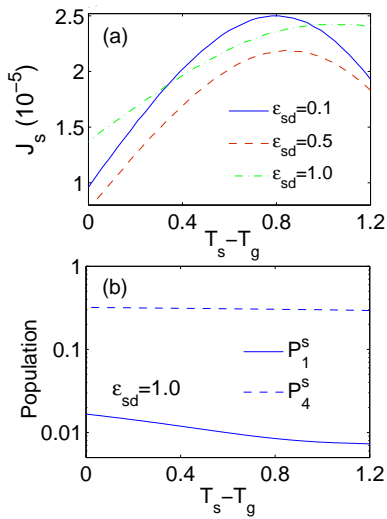}
%\vspace{-2.0cm}
%\end{center}
\caption{(Color online) (a) Influence of $\epsilon_{sd}$ on $J_s$ by tuning $T_g$,
and (b) steady state occupation probabilities $\mathcal{P}^s_1$ and $\mathcal{P}^s_4$
with $\epsilon_{sd}=1.0$, at strong qubit-bath coupling $\alpha_v=\alpha=2.0$.
The other parameters are given by $\Delta=0.1$, $\epsilon_{sg}=\epsilon_{gd}=1$, $T_s=1.6$, $T_d=0.4$ and $\omega_c=10$.}
\label{fig5}
\end{figure}

\section{Influence of $\epsilon_{sd}$ on the NDTC of $J_s$}

It is already known at Fig.~\ref{fig1}(c) that  the NDTC can be exploited with strong qubit-bath interaction,
in which $\epsilon_{sd}$ is assumed zero.
Here, we investigate the influence of the $\epsilon_{sd}$ on $J_s$, shown at Fig.~\ref{fig5}(a).
For weak interaction between source and drain qubits (e.g., $\epsilon_{sd}=0.1$ and $\epsilon_{sd}=0.5$),
the feature of NDTC still appears, though gradually suppressed.
While for strong $\epsilon_{sd}$, NDTC is completely eliminated by tuning $T_g$ in a wide regime.
To exploit the underlying mechanism under the condition of $\epsilon_{vu}=\epsilon~(v{\neq}u)$, we simplify the expression of $J_s$ at Eq.~(\ref{jl1}) as
\begin{eqnarray}
J_s{\approx}4\epsilon(\kappa_{14}\mathcal{P}^s_4-\kappa_{41}\mathcal{P}^s_1),
\end{eqnarray}
at strong qubit-bath coupling, where the term $(\epsilon_{sg}-\epsilon_{sd})(\kappa_{32}\mathcal{P}^s_2-\kappa_{23}\mathcal{P}^s_3)$
is eliminated.
At Fig.~\ref{fig5}(b),
it is found that by modulating $T_g$, $\mathcal{P}^s_4$ is nearly stable,
whereas $\mathcal{P}^s_1$ shows apparent decrease.
Moreover, the transition rates $\kappa_{14}$ and $\kappa_{41}$ are independent on $T_g$.
Hence, the amplitude of $J_s$ increases monotonically.
we conclude that tuning on $\epsilon_{sd}$ is deteriorates to the existence of the NDTC.

\end{document}